# Observation and active control of a collective polariton mode and polaritonic band gap in few-layer WS$_2$ strongly coupled with plasmonic lattices

Wenjing Liu[1], Yuhui Wang[1], Biyuan Zheng[2], Minsoo Hwang[1], Zhurun Ji[1], Gerui Liu[1], Ziwei Li[2], Volker J. Sorger[3], Anlian Pan[2*], and Ritesh Agarwal[1*]

[1]Department of Materials Science and Engineering, University of Pennsylvania, Philadelphia, Pennsylvania 19104, United States

[2]Key Laboratory for Micro-Nano Physics and Technology of Hunan Province, School of Physics and Electronic Science, Hunan University, Changsha, Hunan 410082, P. R. China

[3]Department of Electrical and Computer Engineering, George Washington University 800 22nd St., Science & Engineering Hall, Washington, DC 20052, USA

* Email: anlian.pan@hnu.edu.cn and riteshag@seas.upenn.edu

**Two-dimensional semiconductors host excitons with very large oscillator strengths and binding energies due to significantly reduced carrier screening. Two-dimensional semiconductors integrated with optical cavities are emerging as a promising platform for studying strong light-matter interactions as a route to explore a variety of exotic many-body effects. Here, in few-layered WS$_2$ coupled with plasmonic nanoparticle lattices, we observe the formation of a collective polaritonic mode near the exciton energy and the formation of**

**a complete polariton band gap with energy scale comparable to the exciton-plasmon coupling strength. A coupled oscillator model reveals that the collective mode arises from the cooperative coupling of the excitons to the plasmonic lattice diffraction orders via exciton-exciton interactions. The emergence of the collective mode is accompanied by a superlinear increase of the polariton mode splitting as a function of the square root of the exciton oscillator strength. The presence of these many body effects, which are enhanced in systems which lack bulk polarization, not only allows the formation of a collective mode with periodically varying field profiles, but also further enhances the exciton-plasmon coupling. By integrating the hybrid $WS_2$-plasmonic lattice device with a field-effect transistor, we demonstrate active tuning of the collective mode and the polariton band gap. These systems provide new opportunities for obtaining a deeper and systematic understanding of many body cooperative phenomena in periodic photonic systems and for designing more complex and actively controllable polaritonic devices including switchable polariton lasers, waveguides, and optical logical elements.**

Strong exciton-photon coupling in optical cavities leads to fascinating many body phenomena and has been extensively studied in the past decades[1]. When the coupling strength between the excitonic and photonic states overcomes their decay rates, new half-light-half-matter quasi-particles, i.e., polaritons become the new eignenstates of the system, which combine the unique characteristics of individual uncoupled states. With enhanced nonlinearity, coherence, and reduced effective mass, polaritonic systems enable numerous novel observations and applications including Bose-Einstein condensation [2,3], polariton switches and circuits [4,5], and low threshold polariton lasers.[6,7] Polariton formation and their many attributes have been mostly demonstrated in a variety of low-dimensional semiconductors[8,9] and organic[10] excitonic materials. Recently, strong coupling in mono- and few-layered semiconductors have drawn great interest[11-15] due to their extraordinary excitonic properties resulting from their 2D confinement and reduced Coulomb screening[16], These unique features lead to exceptionally large exciton binding energies (~0.5-1 eV),[17,18] in two-dimensional systems that can enable robust strong coupling up to room temperature. Besides, the reduced screening in the atomically thin structures greatly enhances exciton-exciton and exciton-charge interactions, giving rise to a rich variety of many body effects such as modification of the radiation spectrum and lifetime[19], tightly-bound trion[20] and biexciton formation[21,22], as well as the observation of trion-polaritons[23-25] and polarons[26]. Furthermore, two-dimensional semiconductors also allow effective tuning of their electrical and optical properties via electrostatic doping[23,27,28], hence providing unique opportunities to actively control the systems response, which is critical to realize real-time tunable optical functionalities.

On the other hand, precisely tailored dispersion of the exciton-photon system is critical in studying light-matter interactions in both the strong and weak coupling regimes, and to design photonic devices with novel functionalities. In this context, periodic photonic structures are an attractive platform for creating artificial band structures with anomalous dispersion and optical band gaps that are not readily accessible in natural materials[29]. Photonic bandgap, within which no photonic modes are allowed in periodic dielectric structures, can be utilized to control the optical density of states and create optical cavities with large quality factors and small mode volumes[30], and play an important role in nanophotonic devices[31-34]. Polaritonic systems, including plasmon-, phonon- and exciton-polaritons, are not only harnessed to generate and engineer photonic band structures and band gaps by utilizing the optical properties of active materials[35-37], but also hold promise for active control of polariton dispersion.

Among different photonic crystal systems, periodic arrays of plasmonic nanostructures are of special interest since they support surface lattice resonances (SLRs)[38,39] that arise from the coherent coupling between two distinct types of resonances: localized surface plasmon resonance (LSPR) and the diffractive orders of the periodic lattice. In contrast to dielectric photonic resonances, LSPRs can tightly confine the electric field well below the diffraction limit, leading to enhanced light-matter coupling. On the contrary, lattice diffraction orders have relatively large quality factors and long range spatial coherence. Hence by tuning both resonances independently via geometrical factors, SLRs with precisely tailored fields, quality factors and dispersions can be obtained for different applications including Purcell enhancement of the spontaneous emission[40], strong exciton-plasmon coupling[41], and plasmonic lasers[42].

By exploiting the unique properties of two-dimensional semiconductor excitons coupled to SLRs, we have recently demonstrated Purcell enhancement, Fano resonances and electrically tunable exciton-plasmon polariton coupling between monolayer TMDs and plasmonic lattices.[12,23,40,43] Here, by utilizing the enhanced exciton oscillator strength in few-layered $WS_2$ flakes, we further enhance the exciton-plasmon coupling in the $WS_2$-plasmonic system and report the observation of an emergent collective polaritonic mode near the excitonic energy, in which the electric field profile is extended to the entire lattice, allowing the coherent coupling of distant excitons. The emergence of the collective mode is accompanied by the development of a complete polariton gap, forms between the new collective mode and the upper polariton branch and the gap energy reaches values as high as 45 meV, which is an appreciable fraction of the polariton mode splitting (>100 meV). With the aid of finite difference time domain (FDTD) simulations and a coupled oscillator model, we demonstrate the collective nature of the emerging mode, and show that it further enhances the exciton-plasmon coupling by introducing a large splitting in the exciton absorption spectrum, and a superlinear increase of the mode splitting with the square root of the exciton oscillator strength. Furthermore, we also demonstrate active control of the collective mode and the polariton band gap via electrical gating of the integrated optical device (schematic in Figure 1b).

The reflectance spectrum of as grown few-layered $WS_2$ samples (Figure 1a) were measured at 77 K (see Methods). As the number of layers increases from one to three, in bare samples, the intensity of the differential reflectance dips corresponding to the A exciton increases (Figure 1c) from 0.30, 0.40 to 0.44, which indicates enhanced oscillator strengths in multilayer $WS_2$. Silver plasmonic lattices composed of square shaped Ag nanoparticles with a lattice period of 430 nm were then patterned directly onto the $WS_2$ flakes (Figure 1b, see Methods). The dispersion of the $WS_2$-plasmonic lattice coupled system were measured at 77 K by a home-built angle-resolved reflectance spectrometer[44] (see Methods) with the excitation polarization selected to be perpendicular to $k_{//}$ (TE polarization). As demonstrated in our previous work, the resonances in the system are absorption dominated, hence the polariton modes correspond to the dips (dark region) in the reflectance spectra[43]. In the $WS_2$-plasmonic lattice coupled systems, it was observed that the light-matter coupling strengths also increases with increasing number of layers (Figure 1 d-f). Since the system dispersion arises from the mutual coupling between four resonances: A excitons, LSPR, and the $(\pm 1,0)$ lattice diffraction orders, four polariton branches emerge due to the hybridization of the system's uncoupled resonances (the highest energy polariton branch is above the $\Gamma$ point of the lattice diffraction order and hence does not show up in the spectrum). Indeed, the dispersion of the monolayer $WS_2$-plasmonic lattice hybrid sample exhibits three polariton branches, in accordance to the simple coupled oscillator model (see Supplementary Note 1 for detailed discussion), however with only a small distortion of the dispersion near $\sin\theta = 0.23$.

However interestingly, as the coupling strengths enhances with increasing number of layers, a nearly dispersionless new mode (red dashed line in Figure 1 e and f) appears near the A exciton energy at certain $k_{//}$ in a bilayer sample, which then extends to nearly all angles in the trilayer sample. Moreover, the emergence of the new mode is accompanied by the development of an energy gap (bright region in the reflectance spectra) above the excitonic energy throughout all $k_{//}$, within which no polariton mode exists. The formation of a related energy band gap in polariton dispersion, i.e., a polariton gap, has been observed in intersubband polariton dispersions (THz regime) in the ultrastrong coupling regime[35], and recently numerically predicted in exciton-plasmon coupled nanocavities at optical frequencies[45,46], suggesting that the exciton-plasmon coupling reaches a very strong coupling regime. The observation of the polariton gap in the current system also indicates that the emergent mode near the exciton energy does not simply originate from uncoupled excitons, and instead, it interacts strongly to the existing polariton modes, and significantly modifies the polariton dispersion (details later).

The evolution of the additional mode near the exciton energy and the polariton gap with increasing number of layers are presented in detail as line cut plots of the angle-resolved dispersions in Figure 2 a-c. It can be seen in Figure 2 that the additional mode does not disperse uniformly along different $k_{//}$ in the angle-resolved spectra of monolayer and bilayer WS$_2$-plasmonic lattice samples. The mode first emerges weakly near $\sin\theta = 0.23$ in monolayer samples (Figure 2a), where the lattice diffraction order is in resonance with the A exciton, and hence couples most strongly with the exciton compared with other $k_{//}$. With increasing layer number, this mode extends to lower and higher $k_{//}$, and eventually builds up at nearly all angles (Figure 2c). This finding suggests the important role played by the periodicity of the plasmonic lattice in the formation of the new mode (more discussion later). On the other hand, evidence of a polariton gap can also be observed even in the monolayer sample, albeit with a small gap energy of 18 meV, which increases to 29 meV and 45 meV in bi- and tri- layer samples, respectively, thereby reaching an appreciable fraction of the Rabi splitting value of the exciton-plasmon coupling (~110 meV).

The formation of the additional mode and the polariton gap was studied in more detail with FDTD simulations. The permittivity of the bare WS$_2$ sample with different layers was fitted to the experimentally measured reflectance spectra (Figure 1c) by using a single Lorentzian function:

$$\varepsilon = \varepsilon_{bg} + \frac{f}{\omega_{ex}^2 - \omega^2 - i2\gamma_{ex}\omega} \tag{1}$$

where $\varepsilon_{bg}$ stands for the background permittivity, $\omega_{ex}$ and $\gamma_{ex}$ are the exciton energy and damping, respectively, and $f$ is the oscillator strength, which is proportional to the exciton density, and the excitons were assumed to polarize isotropically in-plane[47]. The possible out-of-plane component of the exciton in multilayer samples is not critical in the current study, leading to only a slight modification in the dispersion when it is incorporated (see Supplementary Figure S7 for details). Although a better fitting might be obtained by a multiple Lorentzian function, we consider here only the A exciton and ignore the influence of other resonances in WS$_2$ for simplicity. Figure 2 d-f present the simulated reflectance spectra at k$_{//}$=0 of the mono-, bi- and tri-layer WS$_2$ coupled with plasmonic lattices, respectively, which are in good accordance with the experimental data, and successfully capture the emergence of the new mode while increasing the thickness of the WS$_2$ layer (see Supplementary Note 2 for the parameters used in the simulations).

To further study the mechanism of the additional mode and polariton gap formation via simulations, we systematically changed the oscillator strength $f$ from 0.2-3 while keeping all other parameters fixed (Figure 3a). While the splitting between the two polariton branches increases with $f$, from $f = 1.5$ onwards, a new mode emerges near the exciton energy at 625 nm which becomes more prominent at high $f$ values. Increasing $f$ also develops an energy gap between the upper polaritons and the additional mode, leading to a significant reflection peak at 622 nm. On the other hand, while $f$ is fixed at 1.5, i.e., the critical value at which the additional mode emerges, the incident angle was swept from $\theta = 0° - 40°$ with TE polarization. Among the five selected angles (Figure 3b), it was observed that the additional mode is most pronounced at $\theta = 10°$, where the lattice diffraction order is in near resonance with the exciton, while at larger angles, this mode vanishes as the lattice diffraction order becomes detuned from the exciton energy. These results, being consistent with experimental observations, indicate that the observed phenomena originate from the enhanced oscillator strengths and is closely correlated to the diffraction order of the lattice. It should be noted that in the current system, increasing the number of layers or increasing $f$ has a similar effect in enhancing the exciton-plasmon coupling. Since, even up to the trilayer sample, the thickness of the $WS_2$ layer (~2.5 nm) is much smaller than the length scale of the plasmonic field, the excitons within the thickness of the $WS_2$ layer can couple coherently to the plasmonic mode. Effectively, it is the exciton oscillator strength per unit area that determines the exciton-plasmon coupling strength in this two-dimensional system.

To further understand the near-field exciton-plasmon interaction of the system especially in the vicinity of the new emergent mode, the polarization dependent absorption in the $WS_2$ layer were extracted from the FDTD simulations. Figure 3c-f and g-j presents the *x*- and *y*- polarized absorption respectively, within a unit cell of the plasmonic lattice patterns under *x*-polarized illumination at 625 nm, where the new mode is located. The absorption–profiles indicate the excitation of the first order LSPR mode of the nanoparticle. At $f = 0.2$, the absorption profile in both polarizations are localized to the plasmonic nanoparticle, since the lattice diffraction order is larger than 200 meV detuned from the wavelength examined. However, interestingly, upon enhancing the exciton oscillator strength $f$, when the additional mode emerges, the material's absorption in the *x*-polarization starts to delocalize from the individual nanoparticle and extends to the entire unit cell and hence the entire lattice. Consequently, periodic high-field regions between the nanoparticles (figure 3i and j) emerge, which are in accordance with the electric field distribution of the TE-polarized SLRs (Supplementary Note 3). Consistently, for inclined incidence (Figure 3 k-n), when the additional mode is strong ($\theta = 10°$), the absorption profile is more delocalized and asymmetric with respect to the y=0 axis, while at high angles the additional mode disappears, the absorption profiles are localized to the individual nanoparticles again. The *x*-polarized electric field profiles at different wavelengths including the $\Gamma$ point, the upper and lower polariton modes and the additional mode were also extracted (Supplementary Note 3), and the extended profiles can only be observed at the $\Gamma$ point and the additional mode, which further confirms our observation. The high polarization anisotropy and the periodic spatial profile of the electric field hence suggests that the additional mode does not result from uncoupled excitons due to the increased number of oscillators in the system, but instead, it originates directly from the formation of a collective mode near the exciton energy coupled to the lattice mode.

The x-polarized exciton absorption spectra from two selected points, A and B (Figure 3g), representing the localized and extended resonances, respectively, were calculated for different $f$ values and incident angles in Figure 3 o-q. Located near the nanoparticle LSPR hot spot, the excitons from point A interact strongly with the LSPR and hence the SLRs. As a result, the exciton absorption spectrum from this point splits from one single peak at $f = 0.2$ to two well separated peaks, which corresponds to the upper and lower polariton branches labeled in Figure 3a, respectively, indicating strong exciton-plasmon coupling. On the other hand, at point B, the absorption spectrum exhibits one main peak near the exciton energy that becomes enhanced but slightly redshifts as the collective mode emerges and develops. These calculations further confirm the relationship between the observed periodic electric field pattern and the collective mode. It should be emphasized that the single peak spectrum at point B does not originate simply from uncoupled excitons; on the contrary, it is a collective resonance from the exciton coupled to the lattice diffraction orders. This point is clearly demonstrated in the angle-resolved absorption spectrum (Figure 3q), when the lattice diffraction order is in near resonance with the excitons ($\theta = 10°$), the absorption peaks are strongly enhanced with two coupled peaks, while only a much weaker peak appears when the lattice diffraction order is detuned at $\theta = 30°, 40°$. As a brief summary, these observations reveal the collective nature of the new mode near the exciton energy, since it arises from the cooperative coupling between the excitons and the periodic lattice and has an extended electromagnetic field profile among the plasmonic lattices that is analogous to the SLRs.

Related phenomena including the formation of collective mode near the exciton energy and polariton band gap opening has been predicted in plasmonic core-shell structures and plasmonic grating coupled with organic semiconductors[45,46,48,49]. In these reports, a collective mode with the electric field distributed uniformly within the excitonic material, i.e., a superradiant mode, was predicted due to electric field mediated exciton-exciton interactions, regardless of the presence of the plasmonic structure and its periodicity. This collective mode can further interact with the polaritons and lead to the formation of polariton gap. Although in our work the detailed experimental observations and FDTD analyses indicate related origins for the new mode and polariton gap, several notable differences are observed. First, the collective mode red shifts with enhancing exciton oscillator strength, in contrast to the blue shift of a pure superradiant mode, implying that the plasmonic structure also plays a role in its formation. Secondly, and significantly, at the wavelength of the collective mode, the electric field does not distribute uniformly within the WS$_2$ layer, on the contrary, it shows a periodic profile. Therefore, besides exciton-exciton interactions, these differences along with the finding that the lattice diffraction order is critical in the collective mode formation highlight the effect of the periodically modulated electric field induced by the plasmonic lattice in our system.

To address the microscopic mechanisms leading to the emergence of the collective mode, we developed a phenomenological coupled oscillator model by solving the equations of motion (EOM) of the oscillators that is modified from a simple EOM model we demonstrated in our previous work (see Supplementary Note 1 and ref. 43). This modified EOM model involves N+3 oscillators: N excitons, LSPR and $(\pm 1, 0)$ lattice diffraction orders. The EOM can hence be written in N+3 linear equations as:

$$\alpha_i^{-1} + 2\sum_j g_{ij} \dot{x}_j = E_0(t) \tag{2}$$

where $i$ and $j$ =1,2,…, N+3 stand for the index of the N+3 dipoles, $\alpha_i$ is proportional to the polarizability of each oscillator, with a form of a Lorentzian resonator: $\alpha_i = \frac{f_i}{\ddot{x}_i + 2\gamma_i \dot{x}_i + \omega_i^2 x_i}$. $x_i$ and $\gamma_i$ are the oscillating amplitude and damping of the resonances, respectively, and $f_i$ denotes the oscillator strength; $g_{ij}$ is the coupling strength between the $i^{th}$ and $j^{th}$ dipoles. $E_0(t)$ denotes the electric field induced by the incident light that excites the LSPR and excitons. In order to account for the periodic spatial modulation of the electric field, in our model we broadly classify the excitons into two types with identical numbers; the excitons that couple strongly to the LSPR, corresponding to point A in Figure 3g, which contribute to the (normal) polariton modes, and the excitons that do not couple to the LSPR, corresponding to point B that contribute to the collective mode, while both types of excitons couple coherently to the lattice diffraction order with identical coupling strengths, $g_{ex-lattice}$. The exciton-exciton interaction is also incorporated in the system with a constant $g_{ex-ex}$, ignoring its dependence on the separation between the excitons for simplicity.[48] The extinction of coupled system was calculated and fitted to the dispersion of Figure 3b (see Supplementary Table S3 for the fitting parameters) and the results are presented in Figure 4a, showing good agreement with the FDTD simulation in both the mode positions and lineshapes. Moreover, the absorption spectra of both types of excitons extracted from the EOM model also match nicely to the absorption spectra (Figure 3o and p), hence suggests that the model correctly captures the important features of the light-matter coupling of the system despite its simplifications. As a comparison, a simple EOM model without exciton-exciton interactions or spatial modifications was also applied and the calculated extinction and absorption spectra are shown in Figure 4a as gray lines, exhibiting no collective modes or polariton gaps in the dispersion (see Supplementary Note 4 for more details). Importantly, the mode splitting in the WS$_2$ absorption spectrum of the polariton modes from the proposed EOM model (and FDTD simulations) is much stronger than from a simple EOM model showing only a broad envelop, suggesting that the development of the collective mode significantly enhances the exciton-plasmon coupling of the system. Such strong enhancement is enabled by the coherent coupling between the excitons contributing to both the collective and polariton modes, via the mutual exciton-exciton interactions and their collective coupling to the lattice diffraction orders.

Due to the large binding energy of the excitons in two-dimensional semiconductors, the Bohr radius of WS$_2$ is estimated to be as small as ~1 nm[50]. As a result, the many body exciton-exciton interaction is greatly enhanced in 2D semiconductors compared with conventional materials, giving rise to the possibility to observe collective phenomena within atomically thin active layers. Besides, according to the FDTD simulations and the proposed EOM model, we emphasize that the plasmonic lattice structure is critical to the development of the collective mode and the polariton gap in our system, allowing such phenomena to occur at much lower exciton density. Indeed, if the spatial modification, i.e., different exciton-LSPR coupling in the equation is removed from the EOM model while keeping the exciton-exciton interaction, only a very weak and dispersive mode can be observed near the excitonic energy, in contrast to our experimental results (see Supplementary Note 4). Therefore, as a result of its sensitivity to the plasmonic lattice geometry, the $WS_2$ multilayer-plasmonic lattice system allows larger feasibility in engineering of the polariton gap width and the strength of the collective mode by tuning the geometrical parameters of the plasmonic lattices (see Supplementary Note 5).

To experimentally study the evolution of the system's dispersion with the exciton oscillator strength, we integrated the $WS_2$-plasmonic lattice system with a field-effect transistor (FET) geometry[23], to actively control the exciton density via carrier injection/depletion (see Methods). For the n-type $WS_2$ flakes applied in our study, increased electron density upon applying positive gate voltages enhances the coulomb screening and leads to reduced exciton oscillator strength and vice versa[23,27]. Figure 4b and d present the reflectance spectra with applied gate voltage from 0 to 80V on $WS_2$ coupled with plasmonic lattices, and Figure 4e shows the tuning of the reflectance spectra of corresponding bare $WS_2$. Upon applying positive gate voltage, the exciton oscillator strength reduces, with a slight red shift of the peak position for less than 1 nm (Figure 4e). As a result, the collective mode diminishes progressively, and eventually disappears (Figure 4d), indicated by yellow arrow), while the system remains in the strong coupling regime. A series of line cuts were taken from the angle-resolved spectra at $\sin\theta = 0.23$, where the collective mode is the strongest, to study the evolution of the dispersion. We also examined the mode splitting between the upper and lower polariton branches from $V_g = 80$ to $0V$ by comparing the experimental result to a simple EOM model (Figure 4g). At $V_g = 80\ V$, the spectra exhibit only two polariton modes, and we fitted this angle-resolved spectrum to the simple EOM model. The mode splitting given by the model at each applied voltage was then calculated based on the exciton oscillator strengths extracted from the bare sample reflectance (Figure 4e), without further adjusting any parameters, resulting in a splitting value that scales linearly with $\sqrt{f}$. At $V_g = 60\ V$, before the formation of the collective mode, the mode splitting of the experimental data follows the trend of the simple EOM model. On the contrary, at $V_g = 40 - 0V$, as the collective mode appears and gradually enhances, the upper polariton blue shifts rapidly away from the collective mode, which eventually leads to the formation of an energy gap of 26 meV at $V_g = 0V$. As a result, the splitting value noticeably deviates from the simple EOM model, displays a threshold-like jump of ~8 meV. This threshold-like behavior suggests the presence of higher order many-body effects, e.g., exciton-exciton interactions, in the exciton-plasmon coupling that leads to the nonlinear increases of the mode splitting. This result also reveals the mechanism of the polariton gap formation, which is due to the repulsive interactions between the collective mode and the upper polaritons. The emergent collective mode interferes destructively with the upper polariton and induces a strong absorption dip in the exciton absorption spectrum, thereby leading to the polariton gap formation. Furthermore, the electrical tuning study demonstrates that the collective mode and the polariton band gap can be actively controlled via electrical gating, which is critical in realizing tunable photonic and polaritonic devices, but not readily accessible in other material systems.

To conclude, few-layered WS$_2$ coupled with plasmonic lattice display enhanced exciton-plasmon coupling with increasing number of layers. Subsequently a collective mode emerges near the exciton energy, along with the formation of an energy gap between the upper polariton and the new collective mode, which is found to arise from the coupling between the excitons mediated by the lattice diffraction orders. Active tuning of the collective mode and the polariton band gap is also realized via electrical gating. These results further our understanding of the exciton-exciton interactions in two-dimensional semiconductors mediated by a periodic field in systems with inadequate carrier screening leading to strong many body and nonlinear effects that can be useful for assembling sensitive and switchable photonic devices.

**Methods**

**Device fabrication:** Few-layered $WS_2$ flakes were grown on 285 nm $SiO_2$/Si substrates via a CVD method with typical flake sizes ranging from 50 to 100 µm. The plasmonic arrays and the source and drain electrodes of the FET gating devices were defined by electron-beam lithography directly onto the $WS_2$ flakes followed by physical vapor deposition of 50 nm Ag and lift-off process. The plasmonic arrays have square lattices with the lattice pitches of 430 nm, and square shaped plasmonic nanoparticles with side lengths varying from 80-110 nm.

**Angle-resolved reflectance measurement:** The angle-resolved reflectance spectra were measured by a homebuilt angle-resolved microscopy setup that was formally described in refs. 12 and 44. White light was focused onto the sample by an objective with NA=0.7 (Olympus, 60X). Reflected light was collected by the same objective, with collimated light focused on the back focal plane (Fourier plane) of the objective. A set of lenses were used to project the back focal plane to the entrance slit of the spectrometer to obtain the angular- and spectrally- resolved image from a CCD camera attached to the spectrometer.

**Figure Captions:**

**Figure 1. Optical properties of bare few-layered WS₂ and WS₂-plasmonic lattice system.** (a) Optical images of as grown few-layered WS$_2$ flakes showing color contrast in mono-, bi- and trilayer samples. (b) Schematic of the multilayer WS$_2$-plasmonic lattice system integrated in field effect transistor device (left panel). Active tuning of the collective mode and the polariton gap in the polariton dispersion can be achieved via applied gate voltage (middle and right panels). In the middle and right panels, blue curves represent upper and lower polariton branches and the red line represents the collective mode. (c) Differential reflectance $((R_{sample} - R_{substrate})/R_{substrate})$ spectrum of the bare and Ag plasmonic lattice of square shaped plasmonic nanoparticles (side lengths 90 nm, lattice pitch $430 \times 430$ nm) coupled WS$_2$ samples at normal incidence. The curves have been offseted for clarity. (d)-(f) Angle-resolved reflectance spectra of mono-, bi- and tri-layer WS$_2$ coupled with Ag plasmonic lattices. White dash lines denote the mode (dip) positions, the yellow arrow indicates the position of the exciton reflectance dip measured from the bare region of the same WS$_2$ flake. The red dashed line indicates the emergent collective mode formation near the exciton energy.

**Figure 2. Line cut plots of the angle-resolved dispersion of plasmonic lattices coupled to WS₂ flakes and their comparison with FDTD simulations.** (a)-(c) Line cut plots extracted from angle resolved dispersion shown in Figure 1(c)-(e) for (a) mono-, (b) bi- and (c) tri-layer WS$_2$ samples coupled to Ag plasmonic lattice in the angle range 0º to 45º. The curves have been offseted for

clarity. FDTD simulation results corresponding to the spectrum at normal incidence (0º) for (d) mono-, (e) bi- and (f) tri-layer WS2 samples coupled to Ag lattice.

**Figure 3. FDTD simulations of optical response of the WS$_2$-plasmonic lattice systems with varying oscillator strength, $f$, and incident angle.** (a) Reflectance spectra with $f$ varying from 0.2 to 3 at normal incidence, $\theta = 0$. Inset: the geometry and coordinates of the WS$_2$-plasmonic lattice system used in the FDTD simulations. (b) Simulated Reflectance spectra with $f = 1.5$ and the incidence angle $\theta$ varying from 0° to 40°. (c)-(f) Simulated exciton absorption profile along $y$-polarization direction (upon $x$-polarized excitation) at $\lambda = 625$ nm, with $f = 0.2$ to 3 and $\theta = 0°$, corresponding to the four reflectance curves in (a). (g)-(j) Simulated exciton absorption profile along $x$-polarization (upon $x$-polarized excitation) at $\lambda = 625$ nm, with $f = 0.2$ to 3 and $\theta = 0°$, corresponding to the four reflectance curves in (a). (k)-(n) Simulated exciton absorption profile along $x$-polarization (upon $x$-polarized excitation) at $\lambda = 625$ nm, with $f = 1.5$ and $\theta = 10°$ to 40°, corresponding to the four reflectance curves in (b). (o)-(p) Simulated exciton absorption spectra selected from point A (from LSPR hot spot region) and B (between two Ag nanoparticles) in (g), respectively, with $f = 0.2$ to 3 and $\theta = 0°$. (q) Simulated exciton absorption spectra with $f = 1.5$ and $\theta = 10°$ to 40°, corresponding to the four reflectance curves in (b). The color scales in (c)-(n) are in logscale.

**Figure 4 Modified equations of motion (EOM) method of a coupled oscillator model and electrical gate tuning of the collective mode and polariton gap of the plasmonic lattice-WS$_2$ system** (a) EOM results of the angle-resolved reflectance and the exciton absorption spectra

corresponding to Figure 3b. (b)-(d) Angle-resolved reflectance spectra of WS$_2$ coupled with plasmonic lattice (lattice pitch 430 nm, nanoparticle side length, 100 nm) at 0, 40 and 80 V applied gate voltage. Only the collective mode and polariton gap are influenced by the applied gate field while the system still remains in the regular strong coupling regime. (e) Reflectance spectrum of bare WS$_2$ from 0 V to 80 V applied gate voltage. (f) Line cuts from the angle-resolved reflectance spectrum of the plasmonic lattice coupled WS$_2$ flakes taken at $\sin\theta = 0.23$ as a function of the applied gate voltage. (g) Comparison between mode splitting between the upper and lower polariton branches measured in experiment and calculated from a simple EOM model. The exciton oscillator strength $f$ is calculated by fitting the bare WS$_2$ reflectance in (e) via FDTD simulation, and the experimental data is extracted from the mode splitting between the upper and lower polariton modes in (f). The mode splitting calculated from the simple EOM model at $V_g = 80\ V$ is fitted to the experimental data, and the rest of the data points are calculated by tuning only the exciton oscillator strengths in the model.

**Acknowledgements**: This work was supported by the US Army Research Office under Grant No. W911NF-12-R-0012-03 and NSF under the NSF 2-DARE program (EFMA-1542879). A.P. is grateful to the National Natural Science Foundation of China (No.51525202), the Aid Program for Science and Technology Innovative Research Team in Higher Educational Institutions of Hunan Province. Nanofabrication and electron microscopy characterization was carried out at the Singh Center for Nanotechnology at the University of Pennsylvania.

**Competing financial interests:** The authors declare no competing financial interests.

**Data availability**: The data that support the findings of this study are available from the authors upon reasonable request.

**Figure 1**

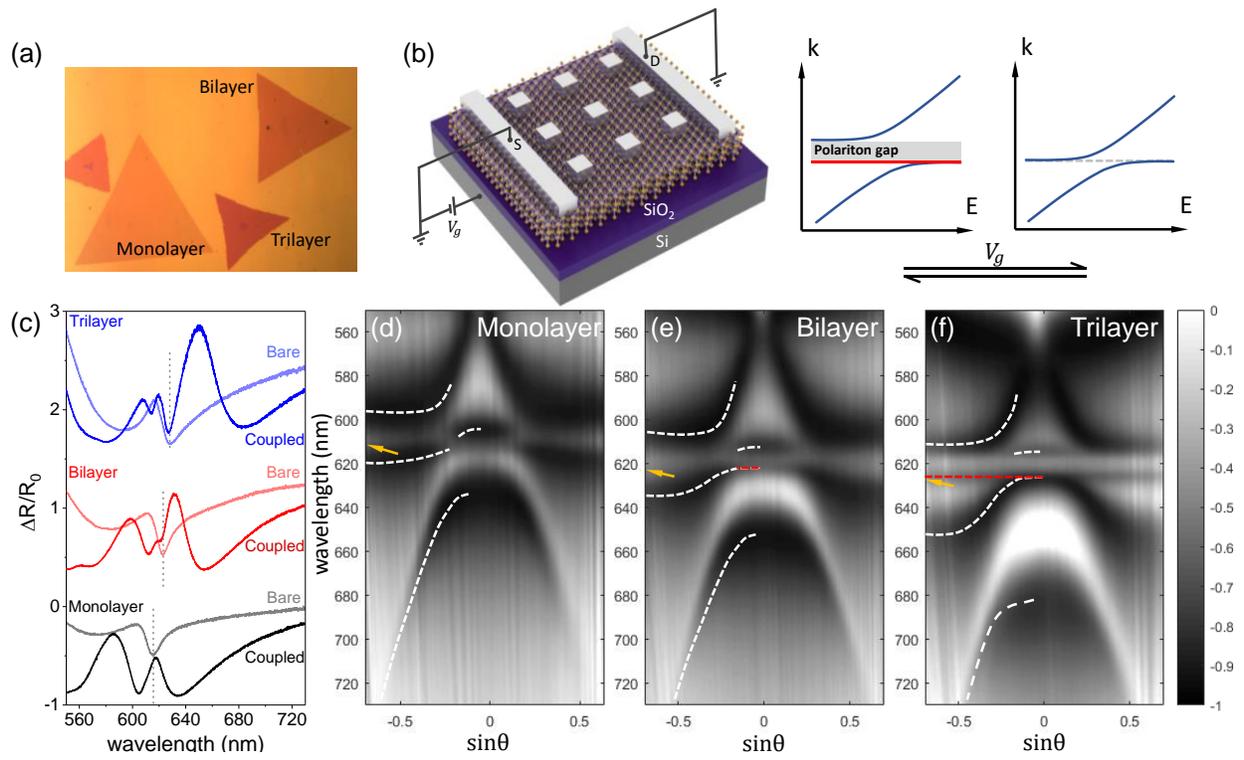

**Figure 2**

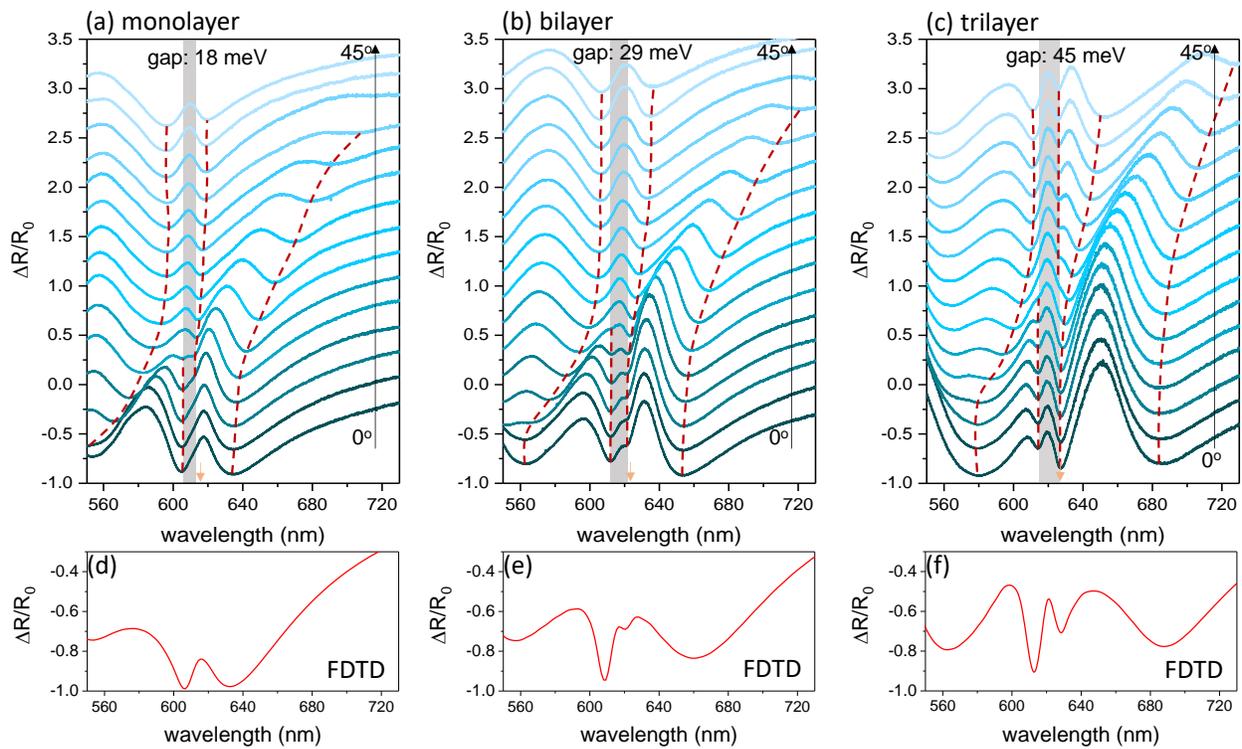

**Figure 3**

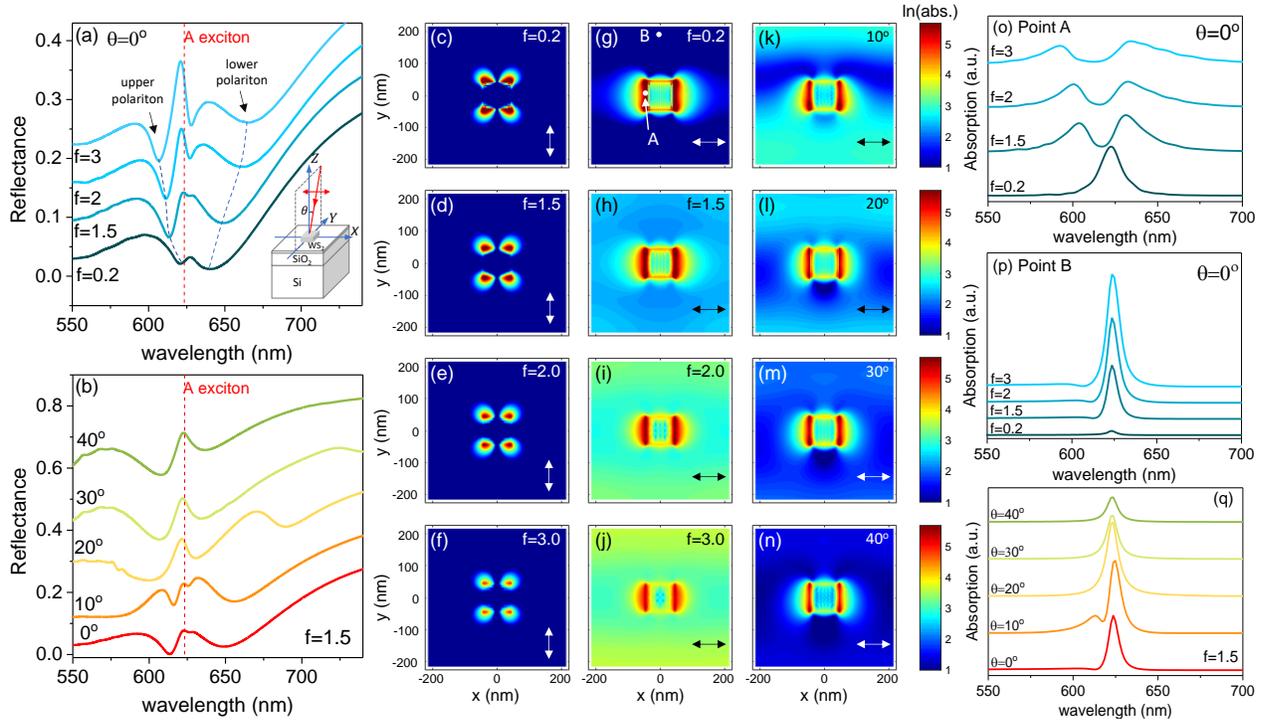

**Figure 4**

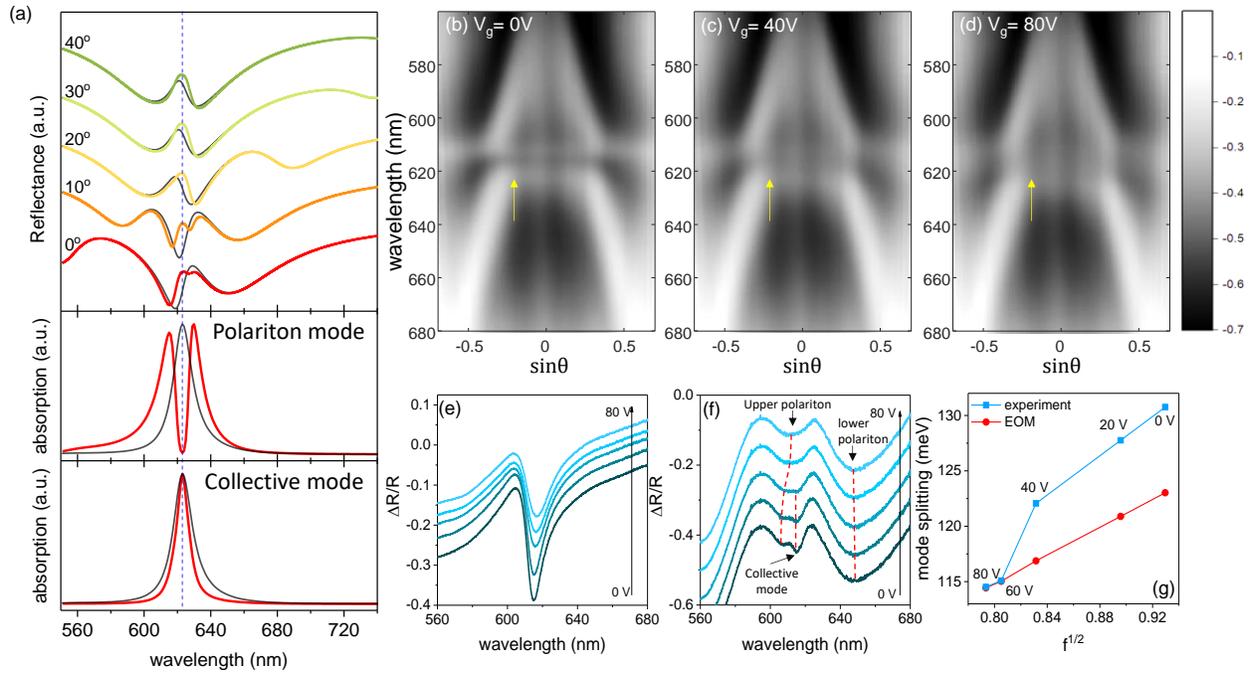